\documentclass[a4paper,12pt]{article}
\usepackage{epsfig}
\usepackage{graphicx}
\usepackage{dcolumn}
\usepackage{bm}
\usepackage{longtable}
\usepackage{amssymb}
\usepackage{amsfonts}
\usepackage[margin=1.0in]{geometry}
\usepackage{hyperref}

\def\dd{\textrm{d}}  				     
\newcommand{\1}{\mbox{$1 \hspace{-1.0mm} {\bf l}$}}  

\makeatletter							  
\@addtoreset{equation}{section}					  
\makeatother							  

\begin{document}

\begin{titlepage}
\begin{flushright}
DFTT 4/2011\\
DISTA-2011
\par\end{flushright}
\vskip 1.5cm
\begin{center}
\textbf{\huge \bf  Fluid Super-Dynamics from\\ \vspace{0.2cm} Black Hole Superpartners}
\textbf{\vspace{2cm}}\\
{\Large L.G.C.~Gentile$ ^{~a,}$\footnote{lorenzo.gentile@gmail.com},$\ $ P.A.~Grassi$ ^{~a,}$\footnote{pgrassi@mfn.unipmn.it}  and A.~Mezzalira$ ^{~b,}$\footnote{mezzalir@to.infn.it}}

\begin{center}
{a) { \it DISTA, Universit\`{a} del Piemonte Orientale,
}}\\
{{ \it via T. Michel, 11, Alessandria, 15120, Italy, }} {{\it INFN- Sezione di Torino}} \\ \vspace{.2cm}
 {b) { \it Dipartimento di Fisica Teorica, Universit\`a di Torino,}}\\
 {{\it via P. Giuria, 1, Torino, 10125, Italy, }}
{{\it INFN
- Sezione di Torino}}
\end{center}

\par\end{center}
\vfill{}

\begin{abstract}
{
       \noindent
Recently the Navier-Stokes equations have been derived from the duality with the black branes in $AdS_5$. The zero modes of black branes are reinterpreted as dynamical degrees of freedom of a conformal fluid on the boundary of $AdS_5$. Here, we derive the corrections to the Navier-Stokes equations due to fermionic zero modes of the black branes. We study only the contributions due to bilinears in the fermionic zero modes in the first order of the parameter expansion. The need of a superextension of the fluid dynamics is a consequence of the 
full $AdS/CFT$ correspondence and yet to be investigated.

}
\end{abstract}
\vfill{}
\vspace{1.5cm}
\end{titlepage}

\vfill
\eject

\tableofcontents
\newpage
\setcounter{footnote}{0}


%


\section{Introduction}
The relativistic Navier-Stokes (NS) equations
\begin{eqnarray}
&&u^\mu \nabla_\mu \rho + \left(\rho + P\right) \nabla_\mu u^\mu  =0 ~,\nonumber \\
&&\left(\rho + P\right) u^\nu \nabla_\nu u^\mu + \left(g^{\mu\nu}+u^\mu u^\nu\right) \nabla_\nu P =0 ~,
\end{eqnarray}
for a conformal fluid can be written in the linearized form in a non-relativistic limit as\footnote{We will just report the results, derivations can be found in \cite{Landau,Wald} and for conformal fluids the precise derivation by Oz et al. \cite{Fouxon:2008tb}.}
\begin{equation}\label{linearized_NS}
3\partial_0 b = \partial_i \beta^i~,  \qquad\qquad \partial_i b = \partial_0 \beta_i~,
\end{equation}
where $b$ is related to the temperature of the black hole by $b=\frac{1}{\pi T}$ and $\beta^{i}$ are the components of the velocity of the fluid. It has been shown in \cite{Bhattacharyya:2008jc,Banerjee:2008th,Erdmenger:2008rm,Bhattacharyya:2008ji,Bhattacharyya:2008kq,Rangamani:2009xk} that equations (\ref{linearized_NS}) are obtained from the Einstein equations as follows: consider the boosted black brane metric\footnote{$\eta_{\mu\nu}$ is the Minkowski metric for the boundary of $AdS_5$.}
\begin{eqnarray}
	\dd s^{2} &=&
	-2u_{\mu}\dd x^{\mu} \dd r
	-r^{2}h\left( b\,r \right) u_{\mu}u_{\nu}\dd x^{\mu}\dd x^{\nu}
	+r^{2}P_{\mu\nu}\dd x^{\mu}\dd x^{\nu}~,
	\label{AdSMinwRang}
\end{eqnarray}
where
\begin{equation}
	u^{\mu}=
	\left(
	\frac{1}{\sqrt{1-\beta^{2}}}~,~
	\frac{\beta^{i}}{\sqrt{1-\beta^{2}}}
	\right)~,
	\qquad
	h\left( r \right)= 1-\frac{1}{r^{4}}~, \qquad P^{\mu\nu}= u^{\mu}u^{\nu}+\eta^{\mu\nu}~.
	\label{adSdesfMR}
\end{equation}
The parameters $b$ and $\beta$ are  taken as local functions of the boundary  and they are reinterpreted as the boundary local degrees of freedom whose dynamics is described by (\ref{linearized_NS}) obtained by inserting solution (\ref{AdSMinwRang}) into Einstein equations.
Contextually, many authors \cite{Compere:2011dx,Bredberg:2011jq,Lysov:2011xx} have further investigated the close relationship between Einstein equations and NS equations, showing how they are related.
In particular, it has been shown how to get the NS equations from different solutions of general relativity (such as  charged and uncharged black branes) and more recently \cite{Bredberg:2011jq,Lysov:2011xx} how to get back to Einstein equation starting from a boundary fluid.
Nevertheless, the complete $AdS/CFT$ correspondence can be only fully established between the supergravity extension of the general relativity and its holographic dual. In our case that would correspond to a supersymmetric extension of Navier-Stokes equations in $4$d.

The aim of the present work is to compute the corrections to the NS equations due to the presence of bulk fermions. In particular, we consider as bulk fermions the superpartners of the zero modes of an uncharged black hole in $AdS_5$, extending the original work by Minwalla et al. \cite{Bhattacharyya:2008jc}: we derive the consistency conditions needed for supergravity equations in terms of some bilinears in fermions. The new dynamical degrees of freedom are implemented in the derivation by taking into account a black hole metric transformed by a generic superisometry of $AdS_5$ space.

In the $5$d case, the Killing vectors are parameterized by $15$ variables which span the isometry group $SO(2,4)$ of $AdS_5$.\footnote{
See \cite{Maldacena:1997re,Aharony:1999ti,Petersen:1999zh} for further details.}
As is well known the black hole partially breaks  those isometries \textit{i.e.} not every $AdS$ Killing vector is a Killing vector for the black hole metric. In particular, taking a static, symmetric black hole in $AdS_5$, there are $7$ $AdS$ Killing vectors which preserve the black hole metric (its $SO(3)$ part) and the variations of the black hole metric $g^{(0)}_{BH}(x^M)$, given by $\mathcal{L}_K(g^{(0)}_{BH})$ with $K$, an $AdS$ Killing vector, depends upon the eight remaining parameters. As in \cite{Bhattacharyya:2008jc} we promote those parameters to local functions on the boundary and by plugging $g^{(1)}_{BH}=g^{(0)}_{BH}(x^M)+\mathcal{L}_K(g^{(0)}_{BH})$ into Einstein equations we derive, at the first order in these parameters and their first derivatives, the modified equations in place of (\ref{linearized_NS}). Our derivation is suitable for both flat and curved boundary. The latter is needed for describing the supersymmetric partner of the black hole zero mode \cite{Behrndt:1998jd,Behrndt:1998ns,Burrington:2004hf}. At first order those constraints coincide, in the large $r$ expansion, with the linearized form of NS equations on a curved or flat space-time. As underlined in \cite{Rangamani:2009xk} the constraints (\ref{linearized_NS}) do not depend on the curvature  $k$ of the boundary.

For what concerns us, we recall that $AdS_{5}$ space has a bigger isometry group which coincides with the supergroup $SU\left( 2,2~|~ 1 \right)$. This means that, besides the bosonic Killing vectors, the $AdS_5$ space admits also $8$ Killing spinors \cite{CastellaniDAuriaFre}, and together with the Killing vectors we can study whether they are also Killing spinors for $g^{(0)}_{BH}$.

In \cite{Behrndt:1998jd,Behrndt:1998ns,Burrington:2004hf}, a general solution for an $AdS_5$ black hole has been constructed depending upon some charges $q_I$ and a non-extremality parameter $\mu$. For simplicity we consider the uncharged case ($q_I=0$) which avoids the analysis of the gauge-field sector.\footnote{To this regard we refer to works \cite{Banerjee:2008th,Erdmenger:2008rm}.} In that case, none of the $AdS$ Killing spinors are Killing spinors for that metric. This is due to the fact that the charges $q_{I}$ must satisfy certain BPS conditions with $\mu$, which are however violated is $q_{I}=0$. The superpartners of the Killing vectors can be read from the violation of the Killing spinor equation.
At this point there are two alternatives. The first one, according to \cite{Bhattacharyya:2008jc}, is promoting the fermionic zero modes to local fermions; then one is forced to study the Rarita-Schwinger equations. Namely, one has to see if certain conditions on the local functions, obtained by promoting the Killing spinors of the $AdS$, lead to dynamical equations for boundary degrees of freedom.
The second alternative is to consider the bilinears in the gravitino fields\footnote{Actually the fermionic bilinears appearing in the metric second variations are $\lambda,\ w_i$ and $N_+$ but only the first one is interesting for our scopes, since the others are from the boundary of $AdS$ and they are not bulk zero modes.} and promoting them to local functions. In that case only the Einstein equations are indeed affected since the we assume that the Rarita-Schwinger equations are preserved at that order and the background has no fermions.
We compute the second variations of the metric $\delta\delta g_{BH}^{(0)}$ which is a linear function of fermionic bilinears and imposing the Einstein equations for this new metric $g_{BH}^{(2)}=g_{BH}^{(1)}+\delta\delta g_{BH}^{(0)}$, in the large $r$ limit, we get
\begin{eqnarray}
	2\left(
	\partial_{i}\beta_{i}
	+3\partial_{0}b
	\right)
	+\sqrt{k}\left(
	w_{i}\partial_{i}\lambda
	+N_{+}\partial_{0}\lambda
	\right)
	&=& 0
	~,\nonumber\\
	2\left(
	\partial_{i}b+\partial_{0}\beta_{i}
	\right)
	-\sqrt{k}\left(
	N_{+}\partial_{i}\lambda+3w_{i}\partial_{0}\lambda
	\right)&=& 0
	~.
	\label{AdScorrecNsIntro}
\end{eqnarray}
The new equations depend explicitly on the curvature $k$ of the boundary and in the limit $k = 0$ they coincide with the original Navier-Stokes equations.

 Obviously these equations must be supplemented by some independent differential equations for $\lambda$. One possibility is to derive it from the conservation of a modified energy-momentum tensor; another one is to derive a complete fluid super-dynamics in such a way that the original NS equations get source terms as in (\ref{AdScorrecNsIntro}).

This paper is organized as follows.
In sec.~\ref{remarks1} we present different metrics for $AdS$ space in generic dimension $p$, in sec~\ref{KV} we show the associated Killing vectors and in sec.~\ref{vielbein} the vielbeins and spin connection.
Sec.~\ref{adsBH} introduces a $AdS_5$ space with black hole; in Sec.~\ref{NS} we briefly review the procedure outlined in  \cite{Bhattacharyya:2008jc,Rangamani:2009xk} and we compute the NS equations in a space with curvature $k$.
In sec.~\ref{bhsuper} we compute the variation of the $AdS_5$ BH metric due to the black hole superpartner: in sec.~\ref{ksAds} solutions of Killing spinors equation for $AdS_5$ is presented, in sec.~\ref{grav} we plug these solutions in the Killing vectors equation for $AdS_5$ with BH, obtaining the variation of gravitinos. In sec.~\ref{metric_correction} through the identities given in sec.~\ref{modes}, the second correction to $AdS_5$ BH metric is computed.
Finally in sec.~\ref{correctionsNS} we derive and analyze the corrections to NS equations.


\section{Remarks on $AdS_{p}$}\label{remarks}

For the purpose of the present work we provide some well-known definitions and conside\-rations about $AdS_p$ metric adapted to our derivation. For the convenience of reader, we describe in some details the Killing vectors and the Killing spinors for $AdS_5$.


\subsection{$AdS_{p}$ Space}\label{remarks1}
We start by considering an $AdS_{p}$ space embedded in ${\mathbb{R}}^{2,p-1}$
\begin{eqnarray}
	x_{0}^{2}+x_{p}^{2}-\sum_{i=1}^{p-1}x_{i}^{2}=R^2
~.
\label{AdSdef}
\end{eqnarray}
with $R$ the radius of $AdS_p$ which from now on we will set to $1$.
In order to obtain a metric for the $AdS$ surface, we can solve equation (\ref{AdSdef}) for $x_p$
\begin{eqnarray}
x_{p}
&=&
\sqrt{
1-x_{0}^{2}+\sum_{i=1}^{p-1}x_{i}^{2}
}
~,
\label{AdSx5}
\end{eqnarray}
and substituting it in the $\mathbb{R} ^{2,p-1}$ metric
\begin{eqnarray}
\dd s^{2}
&=&
\dd x_{0}^{2}
+\dd x_{p}^{2}
-\sum_{i=1}^{p-1}\dd x_{i}^{2}
~,
\label{AdSmetric00}
\end{eqnarray}
we obtain
\begin{eqnarray}
\dd s^{2}
&=&
\left[
\eta_{KL}
+
\frac{
x^{I}x^{J}
}{1-x^{R}\eta_{RS}x^{S}}
\eta_{IK}\eta_{JL}
 \right]
\dd x^{K}\dd x^{L}
~,
\label{AdSmetric2}
\end{eqnarray}
where $\eta_{IJ}=\textrm{diag}\{+,-,\cdots,-\}$ and $\{I,J\}=\{0,1,\cdots,p-1\}$.\footnote{See Appendix~\ref{notation} for a complete list of convention and notation.}
We can choose an alternative parametrization for $AdS_p$
\begin{eqnarray}
x_{0}
=
\cosh \rho \cos \tau
~, \qquad
x_{p}
&=&
\cosh \rho \sin \tau
~, \qquad
x^{i}
=
\sqrt{k}\sinh \rho~ \Omega_{\left(k\right)}^{i} ~,
\label{AdSparam1}
\end{eqnarray}
where $\Omega_{\left(k\right)}^{i}$ represents  the coordinates of a $\left( p-2 \right)$-dimensional space with curvature $k$
\begin{equation}
 \sum_i \Omega_{\left(k\right)}^{i}\Omega_{\left(k\right)}^{i} =\frac{1}{k}
~.
\label{AdSOmega}
\end{equation}
With this choice, metric (\ref{AdSmetric00}) becomes
\begin{eqnarray}
\dd s^{2}
&=&
-\ \dd \rho^{2}
-\sinh^{2}\rho \
\dd\tau^{2}
+
k\cosh^{2}\rho
\ \dd \Omega_{\left(k\right)}^{2}
~,
\label{AdSmetric3}
\end{eqnarray}
As before we can solve equation (\ref{AdSOmega}) for $x_{p-1}$ in order to obtain a metric for this $\left( p-2 \right)$-dimensional hypersurface
\begin{eqnarray}
x_{p-1}
&=&
\sqrt{
\frac{1}{k}-\sum_{i=1}^{p-2}x_{i}^{2}
}
~,
\label{AdS3spaceparam2}
\end{eqnarray}
which leads to
\begin{eqnarray}
\dd \Omega_{\left(k\right)}^{2}
&=&
\left[ \delta_{ij}
+
\frac{x_{i}x_{j}}{\frac{1}{k}-\sum_{i=1}^{p-2}x_{i}^{2}} \right]\dd x^{i}\dd x^{j}
~.
\label{AdSmetricsphere}
\end{eqnarray}
Metric (\ref{AdSmetric3}) is mapped in (\ref{AdSmetric00})  by the following coordinates transformations
\begin{eqnarray}
	\rho
	=
	\textrm{arcsinh} \frac{r}{\sqrt{k}}
	~, \qquad
	\tau
	=
	\sqrt{k}~ t
	~,\qquad
	x_{i}
	= x_{i}
	~.
	\label{AdStrans4}
\end{eqnarray}
It is useful to introduce the metric used in \cite{Bhattacharyya:2008jc,Rangamani:2009xk} (apart from an overall minus sign)
\begin{eqnarray}\label{AdSRMmetric}
\dd s^{2}
&=&
+\left( k + r^{2}\right)\dd t^{2}
-
\frac{1}{k+r^{2}}\dd^{2}r
-
r^{2}\dd \Omega_{\left(k\right)}^{2}
~,
\end{eqnarray}
which can obtained from (\ref{AdSmetric00}) setting
\begin{eqnarray}
	x_{0}
	=
	\sqrt{1+\frac{r^{2}}{k}}\cos\sqrt{k}~t
	~,\qquad
	x_{4}
	=
	r \sqrt{\frac{1}{k}-\sum_{i=1}^{3}x_{i}^{2}}
	~,\qquad
	x_{i}
	= r~ x_{i}
	~.
	\label{AdStrans5}
\end{eqnarray}
To recover the minus sign we Wick rotate all coordinates. Setting for consistency $k \to -k$ we gain the final form we will use for $AdS_p$
\begin{eqnarray}
\dd s^{2}
&=&
-f^2 \dd t^{2}
+
\frac{1}{f^{2}}\dd^{2}r
+
r^{2}\dd \Omega_{\left(k\right)}^{2}
~,
\label{AdSRMmetricTrue}
\end{eqnarray}
where $f= \sqrt{k+r^2}$.


\subsection{Killing Vectors}\label{KV}

Starting from metric (\ref{AdSmetric2}) we can  derive a generic form of $AdS_{p}$ Killing vectors
\begin{eqnarray}
\xi_{I}
&=&
\sqrt{
1-x^{R}\eta_{RS}x^{S}
}\frac{\partial}{\partial x^{I}}
~,
\nonumber\\
\xi^{I}_{\phantom{I}J}
&=&
x^{I}\frac{\partial}{\partial x^{J}}
-x^{J}
\frac{\partial}{\partial x^{I}}
~,
\label{AdSKillVec1}
\end{eqnarray}
that are translations and rotations respectively. Note that
\begin{eqnarray}
	\xi
	&=&
	a_{I}\xi_{I}
	+b_{IJ}\xi_{IJ}
	~,
	\label{AdSgenKilvec}
\end{eqnarray}
 is a generic $AdS$ Killing vector and $a_{I}$, $b_{IJ}$ are the constant parameters of the isometry transformations. Some of them will be promoted to local functions as discussed below.
From these, using the change of coordinates (\ref{AdStrans5})  we write the $AdS_{p}$ Killing vectors components as follows
\begin{eqnarray}
	K^{t}&=& 	
	\frac{1}{f}
	\left(
	r~a_{I}x_{I}\cos~\sqrt{k}t
	+r b_{0I}x_{I}\sin~\sqrt{k}t
	-a_{0}\frac{f}{\sqrt{k}}
	\right)
	~,\nonumber \\
	K^{r}&=&
	\sqrt{k}f
	\left(
	-b_{0I}x_{I}\cos~\sqrt{k}t
	+
	a_{I}x_{I}\sin~\sqrt{k}t
	\right)
	~,\nonumber \\
	K^{i}&=& 	
	-\frac{1}{r}\left[
	\frac{f}{\sqrt{k}}\left( b_{0i} -\frac{1}{2}k x^{i} ~b_{0I}x_{J}\right)\cos~\sqrt{k}t
	+
	\right.
	\nonumber\\&&
	\left.
	+\frac{f}{\sqrt{k}}\left(-a_{i} +k x^{i} ~a_{I}x_{J}\right)\sin~\sqrt{k}t
	 +r~b_{iJ}x_{J} \right]
	~,\label{AdSKilVeci}
\end{eqnarray}
where $b_{iJ}$ is built by a vector $b_{i\left( p-1 \right)}$ and an antisymmetric matrix $b_{ij}$, with $i=\{1,\cdots,p-2\}$ as usual. Moreover, $x_{p-1}$ has to be substituted with $x_{p-1}=\sqrt{\frac{1}{k}-x^{2}}$, where $x^{2}=\sum_{i=1}^{p-2}x_{i}^{2}$. Notice that in $AdS_5$ case we find $15$ free parameters as expected, being $15$ the dimension of conformal group in $4$-dimensions $SO(2,4)$.


\subsection{Vielbeins and Spin Connection}\label{vielbein}

Solving vielbeins equation for metric (\ref{AdSmetric2}) we find
\begin{eqnarray}
	e^{I}
	&=&
	\dd x^{I}+\frac{1-\sqrt{1-x^{2}}}{x^{2}\sqrt{1-x^{2}}}x^{I} x\cdot \dd x
~.
\label{AdSvielbeins}
\end{eqnarray}
Imposing null torsion, we work out the spin connection
\begin{eqnarray}
	\omega^{I}_{\phantom{I}J}
	&=&
	\frac{1-\sqrt{1-x^{2}}}{x^{2}}
	\left( x^{I}\dd x_{J}-x_{J}\dd x^{I} \right)
~.
\label{adSspinConn}
\end{eqnarray}
Vielbeins for metric (\ref{AdSRMmetric}) can be derived from (\ref{AdSvielbeins}) by a coordinate transformations or by direct computation
\begin{eqnarray}
	e^{0}
	=
	f\dd t
	~,\qquad
	e^{1}
	=
	\frac{1}{f}\dd r
	~,\qquad
	e^{i}
	=
	r\left( 	\dd x^{i}+\frac{1-\sqrt{1-k x^{2}}}{x^{2}\sqrt{1-k x^{2}}}x^{i} x_j \dd x^j \right)
	~,
	\label{AdSvielbainsTrue}
\end{eqnarray}
notice that vielbeins indices are flat, that is
\begin{eqnarray}
e^{i}e^{j}\eta_{ij}&=& g_{\mu\nu}\dd x^{\mu}\dd x^{\nu}
~.\label{AdSdefviel}
\end{eqnarray}
This time the spin connection is given by
\begin{eqnarray}
	\omega^{i}_{\phantom{i}0}= 	0
	~,\qquad&&
		\omega^{0}_{\phantom{0}1}= r~\dd t
	~,\nonumber \\
	\omega^{i}_{\phantom{i}1}= \frac{f}{r}e^{i}
	~,\qquad&&
	\omega^{i}_{\phantom{i}j}
	=
	\frac{1-\sqrt{1-k x^{2}}}{x^{2}}
	\left( x^{i}\dd x_{j}-x_{j}\dd x^{i} \right)
~.
	\label{AdSspinconnTrue}
\end{eqnarray}


\section{Bosonic Construction}\label{bosonic}


\subsection{$AdS_{5}$ with Black Hole}\label{adsBH}

In order to have a black hole in $AdS$ we modify our metric (\ref{AdSRMmetricTrue}) as in \cite{Behrndt:1998jd} introducing a \textit{non-extremality} parameter $\mu$. The metric in presence of black hole reads
\begin{eqnarray}
\dd s^{2}
&=&
-\left( f^{2}+\frac{\mu}{r^{2}}\right)\dd t^{2}
+
\frac{1}{f^{2}+\frac{\mu}{r^{2}}}\dd r^{2}
+
r^{2}\dd \Omega_{\left(k\right)}^{2}
~.
\label{AdSBHmetric}
\end{eqnarray}
Working out vielbeins we find
\begin{eqnarray}
e^{0}
&=&
\sqrt{f^{2}+\frac{\mu}{r^{2}}}
~\dd t
~,\nonumber\\
e^{1}
&=&
\frac{1}{\sqrt{f^{2}+\frac{\mu}{r^{2}}}}
~\dd r
~,\nonumber\\
e^{i}
&=&
r\left( 	\dd x^{i}+\frac{1-\sqrt{1-k x^{2}}}{x^{2}\sqrt{1-k x^{2}}}x^{i} x_j \dd x^j \right)
	~.
\label{AdSvielbeinsBH}
\end{eqnarray}
Again, with these vielbeins we compute the spin connection components finding
\begin{eqnarray}
	\omega^{i}_{\phantom{i}0}= 	0
	~,\qquad &&
	\omega^{0}_{\phantom{0}1}= \left( r-\frac{\mu}{r^{3}} \right)~\dd t
	~,\nonumber\\
	\omega^{i}_{\phantom{i}1}=\frac{\sqrt{f^{2}+\frac{\mu}{r^{2}}}}{r}e^{i}
	~,\qquad &&
	\omega^{i}_{\phantom{i}j}
	=
	\frac{1-\sqrt{1-k x^{2}}}{x^{2}}
	\left( x^{i}\dd x_{j}-x_{j}\dd x^{i} \right)
~.
	\label{AdSBHspinconn}
\end{eqnarray}


\subsection{Navier-Stokes Equations}\label{NS}

Here we sum up the technique developed in \cite{Bhattacharyya:2008jc,Rangamani:2009xk}. Let $g^{(0)}_{BH}$ be the metric of a spherical, symmetric, planar, uncharged black hole in $5$d. After deriving Killing vectors $K^M$ for $AdS_5$ (see previous section)  we compute Lie derivative of the metric $\mathcal{L}_K\left(g^{(0)}_{BH}\right)$. As said, this depends only on Killing vector parameters $\{\chi\}$ which break  $g^{(0)}_{BH}$ isometries. The modified metric is then
\begin{equation}
	 g_{BH}^{(1)}\left(x^M\right)=g^{(0)}_{BH}(x^M)+\mathcal{L}_K\left(g^{(0)}_{BH}\right)\left(x^M\right) ~.
\end{equation}
The metric $g_{BH}^{(1)}$ with constant parameters $\{\chi\}$ is still a solution of Einstein equations. Now, we promote $\{\chi\}$ to local functions of the boundary coordinates $\varepsilon x^\mu$, where $\varepsilon$ is a formal counting parameter of the number of derivatives. After that, the metric satisfies Einstein equations when some constraints are imposed on the set $\{\chi\}$.
The constraints in the large $r$ expansion and in a neighborhood of the origin, can be identified with non-relativistic, linearized NS equations.

Applying the above procedure to metric (\ref{AdSBHmetric}) we find
\begin{eqnarray}
	A\left( r,k,\mu \right)~\partial_{i}a_{i}-B\left( r,k,\mu \right)~3\partial_{0}b_{04}&=& 0
	~,\nonumber\\
	C\left( r,k,\mu \right)~	\partial_{i}b_{04}-D\left( r,k,\mu \right)~\partial_{0}a_{i}&=& 0
	~,
\label{AdSNSeqstot}
\end{eqnarray}
where we have defined
\begin{eqnarray}
	A\left( r,k,\mu \right)&=& \mu
	\left( 4r^{6}+4k^{2}r^{2}-\mu k+kr^{4} \right)
	~,
	\nonumber\\
	B\left( r,k,\mu \right)&=&
	\mu	\left(
	4r^{6}+18 k^{2}r^{2}-7k r^{4}
	\right)
	~,
	\nonumber\\
	C\left( r,k,\mu \right)&=&
	\mu\left(
	4r^{4}\left( -\mu+r^{4} \right)+
	2k^{3}r^{2}
	-2k^{2}\mu
	+8k^{2}r^{4}-5\mu k r^{2}+10k r^{6}
	\right)
	~,
	\nonumber\\
	D\left( r,k,\mu \right)&=&
	\mu	\left(
	4r^{4}\left( -\mu+r^{4} \right)
	+3k^{2}r^{4}
	-2\mu k r^{2}+7k r^{6}
	\right)
	~.
	\label{AdSNSdefs}
\end{eqnarray}
We have used the notation explained after eq.~(\ref{AdSKilVeci}) to denote the parameters $a_{i}$  and $b_{04}$ as the zero mode. Now, these parameters are local functions of the boundary.
Setting $k=0$, eqs. (\ref{AdSNSeqstot}) become the usual linearized NS equations for a non-viscous fluid as seen in \cite{Bhattacharyya:2008jc,Rangamani:2009xk}.
It is interesting to analyse (\ref{AdSNSeqstot}) on the $AdS_{5}$ boundary $\mathbb{R}^{+}\times \Omega_{\left(k\right)}$ obtained for $r\rightarrow\infty$
\begin{eqnarray}
	\partial_{i}a_{i}
	-
	3\partial_{0}b_{04}&=& 0
	~,\nonumber\\
	\partial_{i}b_{04}
	-
	\partial_{0}a_{i}&=& 0
	~,
\label{AdSNSeqstotLim}
\end{eqnarray}
which once again are the linearized NS equation for ideal fluids ({\it i.e.} the Euler equations for hydrodynamics). Note that there is no dependence  on $k$.

The presence of a curved $3$-dimensional space seems to perturb the behavior of the ideal fluid only for finite $r$. For $r\rightarrow\infty$ the fluid described by the parameters of the $AdS_{5}$ broken isometries is ideal for every $k$ (see \cite{Rangamani:2009xk} and references therein).


\section{Black Holes Superpartner}\label{bhsuper}


\subsection{Killing Spinors for $AdS_{5}$}\label{ksAds}

Let us now introduce Dirac gamma matrices in an arbitrary dimension. As is known, they form the Clifford algebra
\begin{equation}
\left\{ \Gamma_a , \Gamma_b\right\} = 2 \eta_{ab} ~.
\end{equation}
This algebra has an analogous in curved space, obtained introducing vielbeins
\begin{equation}
\left\{\Gamma_a e^a_\mu , \Gamma_b e^b_\nu \right\} = 2 \eta_{ab}e^a_\mu e^b_\nu = 2g_{\mu\nu} ~.
\end{equation}
This observation allows us to write an analogous of Killing equation for  spinors
\begin{eqnarray}
	\dd \epsilon
	+\frac{1}{4}\omega^{ab}\Gamma_{ab}
	~\epsilon
	+
	\frac{1}{2}e^{a}\Gamma_{a}
	~\epsilon
	&=& 0
	~,
	\label{AdSKillSpineq}
\end{eqnarray}
where we have defined
\begin{equation}
\frac{1}{2}\left[\Gamma_a , \Gamma_b \right] = \Gamma_{ab}~.
\end{equation}
Writing equation (\ref{AdSKillSpineq}) in components we find
\begin{eqnarray}
	0&=&	
	\partial_{1}\epsilon
	+\frac{1}{2f}\Gamma_{1}\epsilon	
	~,
	\nonumber\\
	0&=&
\partial_{0}\epsilon
	+
	\frac{1}{2}
	\Gamma_{0}
	\left(
	r\Gamma_{1}+f
	\right)
	\epsilon	
	~,
	\nonumber\\
	0&=&
\nabla_{a}\epsilon
	+\frac{1}{2}
	e^{b}_{\phantom{j}a}
	\Gamma_{b}
	\left[
	\frac{1}{r}f
	~\Gamma_{1}
	+1
	\right]\epsilon
	~,
	\label{AdSKilSpinEqs2}
\end{eqnarray}
where $	\nabla_{i}\epsilon =	 \partial_{i}\epsilon+\frac{1}{4}\omega^{jk}_{\phantom{jk}i}\Gamma_{jk}\epsilon$.
As explained in \cite{Burrington:2004hf}, we decompose the Hilbert space into two subspaces: one related to $r,t$ directions and the other corresponding to $\Omega_{\left(k\right)}$.
 To find solutions to (\ref{AdSKilSpinEqs2}) we decompose $5$-dimensional Dirac gamma matrices in
\begin{eqnarray}
\Gamma_{0}=i\sigma_{2}\times \1
~,\quad
\Gamma_{1}=\sigma_{1}\times \1
~,\quad
\Gamma_{i}=\sigma_{3}\times\sigma_{i}
~,
\label{AdSdefMatrices}
\end{eqnarray}
where all indices are flat. With these definitions, the adjoint spinor is defined as $\bar\epsilon=\epsilon^{\dagger}\Gamma_{0}$.
Using the above factorization the solutions to (\ref{AdSKilSpinEqs2}) can be written as\footnote{These solutions are the generalization for a generic $k$ of the ones found in \cite{Burrington:2004hf}.}
\begin{eqnarray}
	\epsilon_{\pm+} = \left(\sqrt{f+\sqrt{k}}-\sqrt{f-\sqrt{k}}\sigma_1\right)\left(1\pm\sigma_2\right)\varepsilon_{0\pm} \times \eta_{\pm}~,
	\label{AdSsolutionspinor}
\end{eqnarray}
where $\varepsilon_{0\pm}$ are $2$-dimensional Majorana (real) spinors and $\eta_{\pm}$ are spinors defined for the $3$-dimensional $r$-normalized space $\Omega_{\left(k\right)}$.
Notice that, by construction, (\ref{AdSsolutionspinor}) satisfy the projector relations
\begin{equation}
P_{\pm+}\epsilon_{\pm+}=0~,
\end{equation}
where
\begin{eqnarray}
	P_{\zeta\bar\zeta}
	&=&
	\frac{1}{2}
	\left[
	1
	+
	\frac{1}{f}
	\left( i~\zeta\sqrt{k}\Gamma_{0}+\bar\zeta~r\Gamma_{1} \right)
	\right]
	~,\quad\quad
	\textrm{with } \zeta,~\bar\zeta = \pm 1
	~.
	\label{AdSprojectors0}
\end{eqnarray}
The number of independent fermions is established by observing that the Dirac spinor in 5d is decomposed into 2 real Majorana and 2 complex components of a
$SU(2)$-spinors (for $k>0$) or 2 complex components of $SU(1,1)$-spinor for $k<0$.
From now on we will drop subscripts $\pm +$, using for convenience, $\epsilon_{++}$ since both $\epsilon_{++}$ and $\epsilon_{-+}$ generates the same gravitino variations (see next section).


\subsection{Gravitino Variations}
\label{grav}

We decompose in components the Killing equation (\ref{AdSKillSpineq}) using the vielbeins and spin-connection. We obtain
\begin{eqnarray}
	\delta\psi_{r}&=&
		\partial_{r}\epsilon
		+\frac{1}{2\sqrt{f^{2}+\frac{\mu}{r^{2}}}}
	\Gamma_{1}\epsilon
	~,
	\nonumber\\
	\delta\psi_{t}&=&
		\partial_{t}\epsilon
	+\frac{1}{2}\Gamma_{0}
	\left[
	\left(
	r-\frac{\mu}{r^{3}}
	\right)\Gamma_{1}
	+
	\sqrt{f^{2}+\frac{\mu}{r^{2}}}~
	\right]\epsilon
	~,
	\nonumber\\
	\delta\psi_{i}&=&
		\nabla_{i}\epsilon
	+\frac{1}{2}
	e_{i}^{\phantom{a}b}
	\Gamma_{b}
	\left[
	\frac{1}{r}\sqrt{f^{2}+\frac{\mu}{r^{2}}}
	~\Gamma_{1}
	+1
	\right]~\epsilon
	~,
	\label{AdSBHKillspinEq}
\end{eqnarray}
then, using (\ref{AdSKilSpinEqs2}) and   expanding over $\mu=0$ we have
\begin{eqnarray}
	\delta\psi_{r}&=&
	-\frac{1}{4 f^{3}} \frac{\mu}{r^{2}} \Gamma_{r} \epsilon
	~,
	\nonumber\\
	\delta\psi_{t}&=&
	\frac{1}{2}\Gamma_{t}\left(
	-\frac{1}{r}\Gamma_{r}+\frac{1}{2f}
	 \right)\frac{\mu}{r^{2}}\epsilon
	~,
	\nonumber\\
	\delta\psi_{i}&=&
	 \frac{r}{4f}\hat\Gamma_{i}\Gamma_{r}\frac{\mu}{r^{2}}\epsilon
	~,
\label{AdSVariationgravitinos}
\end{eqnarray}
where we defined the Dirac matrices  for the unit $3$-dimensional $\Omega_{\left(k\right)}$ space $\hat\sigma_{i}$
\begin{eqnarray}
	\hat\Gamma_{i}
	&=&
	\sigma_{3}\times\hat\sigma_{i}
	=\sigma_{3}\times\sigma_{j}\frac{e^{j}_{\phantom{i}i}}{r}
	~.
	\label{AdSdefhatgamma}
\end{eqnarray}


\subsection{Zero Mode Identities}\label{modes}
We now sum up some useful identities that will come in handy to compute the metric second variations
\begin{eqnarray}
\bar\epsilon\epsilon
=
-8\sqrt{k}~\lambda~N
~,&&\quad\quad
\bar\epsilon\Gamma_{0}\epsilon
=
-8if\sqrt{k}~\lambda~N
~,\nonumber\\
\bar\epsilon\Gamma_{1}\epsilon
=
0
~ ,&&\quad\quad
\bar\epsilon\Gamma_{1}\Gamma_{0}\epsilon
=
-8ir\sqrt{k}~\lambda~N
~,\nonumber\\
\bar\epsilon\hat\Gamma_{i}\Gamma_{0}\Gamma_{1}\epsilon
=
-8\sqrt{k}~\lambda\hat K_{i}
~, &&\quad\quad
\bar\epsilon\hat\Gamma_{i}\Gamma_{0}\epsilon
=
0
~,\nonumber\\
\bar\epsilon\hat\Gamma_{i}\epsilon
=
8ir~\lambda~\hat K_{i}
~,&&\quad\quad
\bar\epsilon\hat\Gamma_{i}\Gamma_{1}\epsilon
=
-8if~\lambda~\hat K_{i}
~,\label{AdSBIL1}
\end{eqnarray}
where we defined
\begin{eqnarray}
\lambda=\varepsilon_{-}\varepsilon_{+}
~,\quad\quad
N=
\eta^{\dagger}\eta
~,\quad\quad
\hat K_{i}
=
\eta^{\dagger} \frac{\sigma_{j}e^{j}_{i}}{r} \eta
~,\label{AdSBILBil}
\end{eqnarray}
with $e^{j}_{i}r^{-1}$ and  $\hat K_{i}$ respectively vielbeins  and   Killing vectors for the unit $3$-dimensional space parameterized by $k$.
\begin{eqnarray}
\hat K^{i}
&=&
w_{jk}\left[
\varepsilon^{j k i}
\sqrt{1-k\left(\sum_{i=1}^{3} x_{i}^{2} \right)}
+\sqrt{k}\left(
x^{j}\delta^{k i}
-x^{k}\delta^{j i}
 \right)
 \right]
~,
\label{AdSUnitsphereKV}
\end{eqnarray}
where $w_{j k}$ is a $3$-dimensional antisymmetric matrix of parameters.

\subsection{Correction to Metric} \label{metric_correction}

Since we are only interested in the correction to the metric, we assume that the fermionic zero modes do not depend on the boundary coordinates and we compute the second order variations of the metric. Then we allow the bilinear $\lambda$ to become $x^{\mu}$-dependent, we insert these variations in the Einstein equations and we apply again the procedure outlined above.
Using these results we obtain
\begin{eqnarray}
\delta\delta g_{tt}
&=&
f^{2} \frac{2\mu\sqrt{k}}{r^{2}f^{2}}~\lambda~ N~,
\nonumber\\
\delta\delta g_{rr}
&=&
 \frac{1}{f^{2}}\frac{2\mu\sqrt{k}}{r^{2}f^{2}}~\lambda~ N
~,\nonumber\\
\delta\delta g_{ti}
&=&
\frac{3\mu\sqrt{k}}{r^{2}}~\lambda~\hat K_{i}
~,\nonumber\\
\delta\delta g_{ri}
&=&
\delta\delta g_{rt}
=
\delta\delta g_{ij}=0
~.\label{AdSdeltaduemetric}
\end{eqnarray}
The last variations are absent because of the remaining symmetries of the solutions. The other variations are proportional to the fermionic bilinears. The product of the two bilinears is coming from the usual factorization of the spinors. Obviously, they are proportional to the parameter $\mu$ responsible for the the BH metric.


\section{Corrections to Navier-Stokes Equations}\label{correctionsNS}

We compute the simultaneous variation under Killing vectors and Killing spinors of the metric (\ref{AdSBHmetric}) expanded around $\mu=0$. Imposing that the Einstein equations hold for the obtained metric with local parameters we get the following constraints, which can be seen as corrected NS equations.
\begin{eqnarray}
0&=&
2r^{3}\left( \partial_{i}a_{i}-3 \partial_{0}b_{04} \right)
+2k r\left( \partial_{i}a_{i}-\frac{9}{4} \partial_{0}b_{04} \right)
+\nonumber\\&&
-\sqrt{k\left( k+r^{2} \right)}
\left(
\varepsilon_{ijk}\partial_{i}\lambda w_{jk}
\left(
3k+2r^{2}
 \right)
-r^{2}N\partial_{0}\lambda
 \right)
~,
\label{AdScorrectiontoNS1}
\end{eqnarray}
and
\begin{eqnarray}
0&=&
-2k^{2}\partial_{i}b_{04}
+
4r^{4}\left(-\partial_{i}b_{04}+\partial_{0}a_{i} \right)
+3kr^{2}
\left(
-2\partial_{i}b_{04}
+\partial_{0}a_{i}
 \right)
+
\nonumber\\&&
+2r\left(
r^{2} +k\right)\sqrt{k\left( k+r^{2} \right)}
\left(
N\partial_{i}\lambda-6\partial_{0}\lambda\varepsilon_{ijk}w_{jk}
 \right)
~.
\label{AdScorrectiontoNS2}
\end{eqnarray}
In the limit $r\rightarrow\infty$ these became
\begin{eqnarray}
	2\left(
	\partial_{i}a_{i}
	-3\partial_{0}b_{04}
	\right)
	+\sqrt{k}\left(
	w_{i}\partial_{i}\lambda
	+N\partial_{0}\lambda
	\right)
	&=& 0
	~,\nonumber\\
	2\left(\partial_{0}a_{i}-
	\partial_{i}b_{04}
	\right)
	-\sqrt{k}\left(
	N\partial_{i}\lambda+3w_{i}\partial_{0}\lambda
	\right)&=& 0
	~.
	\label{AdScorrecNs}
\end{eqnarray}
where $w_{i}=\varepsilon_{ijk}w_{jk}$. Setting to constant the parameters of the Killing vectors, we get equations for $\lambda$ only.
The compatibility condition is
\begin{eqnarray}
	N^{2}&=& 3w_{i}w_{i}
	~.
	\label{AdScompeq}
\end{eqnarray}
The new equations depend explicitly on the curvature $k$ of the boundary and in the limit $k = 0$ they coincide with the original Navier-Stokes equations. By computing the additional derivatives
and forming linear combinations, one finds compatibility condition (\ref{AdScompeq}).
However, this is guaranteed by the Fierz identity among $SO(3)$ spinors appearing in $w_i$ and $N$.

The corrections due to the other bilinears $N$ and $K^i$ to the NS equations can be
also be computed along the same line. In particular, it would be rather interesting to see
whether they also affect the NS equations. Another missing sector of our analysis is the
complete study of the supergravity equations for a more general BH solution depending of the
charges. In that case, the extremality conditions implies that some of the superisometries are
preserved and it would be interesting to study the modifications to the NS equations.


\section*{Conclusions}\label{conclusion}
\addcontentsline{toc}{section}{Conclusions}

We present a computation of the fermionic corrections to Navier-Stokes equations based
on the gravity/fluid-dynamics correspondence. We promote the fermionic zero modes to
local functions of the boundary coordinates and we derive the constraints on those
functions due to Einstein equations. We simplify the analysis studying only the corrections
to the bilinears in the fermionic zero modes and we intend to provide a complete analysis
in future papers.

The present work opens up to several new generalizations: 1) Instead of taking as local
functions only the bilinears, one can promote the fermions themselves to be local functions
of the boundary coordinates. In that case, one needs to study the constraints coming from
the Rarita-Schwinger equations in the same spirit as done in work on Minwalla {\it et al}.
Apparently, nobody performs this analysis and we think that it might be very instructive and
useful. 2) What are the supersymmetric Navier-Stokes? A complete analysis of the
supersymmetrization of NS eqs. is missing and, more importantly, their couplings with
supergravity. We hope to report soon on this subject.

\section*{Acknowledgments}
We thank G. Policastro and L. Sommovigo for useful comments and discussions.

\vskip 1cm
\appendix
\addcontentsline{toc}{section}{Appendices}

\noindent {\Large \bf Appendices}
\section{Notations and Conventions} \label{notation}
Here you will find a list of symbols and conventions used throughout this paper.\\
Lower case latin indices run from $1$ to $3$ or from $3$ to $5$ while upper case ones run from $0$ to $p$. Einstein's sum convention is in use as well.
\begin{longtable}{p{0.5\textwidth}p{0.5\textwidth}}
\Large{Expression} & \Large{Meaning} \\
\hline
& \\
$\eta_{\mu\nu} = diag(1,-1,-1,-1,-1)$  &  Minkowski metric \\
$g_{\mu \nu}$ & Curved metric\\
$T$ & Temperature \\
$b$ &  $b= \frac{1}{\pi T}$\\
$\beta$ & Boost or fluid velocity\\
$\rho$ & Energy density\\
$P$ & Pressure \\
$u^\mu$ & Velocity field \\
$\eta_{\pm}$ & Killing spinors for $\Omega_{\left( k \right)}$ \\
$\mathcal{L}_X$ & Lie derivative respect to vector X \\
$\Omega_{\left(k\right)}$ & Boundary coordinates $\sum_i \Omega_i^2 = 1$\\
$k$ & Parameter of space curvature\\
$f$ & $\sqrt{k+r^2}$ \\
$\Gamma$ & Dirac's matrix in an arbitrary dimension\\
$\mu$ & Non-extremality parameter
\end{longtable}

\section{Navier-Stokes Equations on the Sphere}
As a test of our results we compute the NS equations on the sphere through the covariant conservation of the stress tensor. We start with the metric of a unitary $\mathcal{S}^3$ sphere embedded in $\mathbb{R}^{1,3}$ parameterized by cartesian coordinates\footnote{Note that in this section \textit{latin} indices are contracted with $\delta_{ij}$.}
\begin{equation} \label{rxsphere}
ds_\Omega^2=-dt^2+ d\Omega_3^2
\end{equation}
where
\begin{eqnarray} \label{cartesian_sphere}
d\Omega_3^2 &=&\left(dx^i\right)^2+\frac{x_i x_j dx^i dx^j}{1-\sum_{i=1}^3\left(x_i\right)^2}
\end{eqnarray}
The stress tensor of a perfect conformal fluid is (see \cite{Fouxon:2008tb,Rangamani:2009xk} for an explicit construction)
\begin{equation}
T^{\mu\nu} = \frac{1}{b^4}\left(g^{\mu\nu}+4u^\mu u^\nu\right)
\end{equation}
We choose the normalization for the velocity field $u^\mu u_\mu = -1$ (notice that indices are raised and lowered with metric \ref{rxsphere}). The field that does the trick is
\begin{eqnarray}
u^0&=& \sqrt{\frac{1-x \cdot x + \left(x \cdot \beta \right)^2}
{\left(1-x\cdot x\right) \left(1-\beta \cdot \beta \right)}} \\
u^i &=& \frac{\beta^i}{\sqrt{1-\beta \cdot \beta}}
\end{eqnarray}
with $\cdot$ the scalar product for the $3$-vector.
At zero order the stress tensor is in a neighborhood of the origin, as expected, covariantly conserved. If we promote $\beta_i$ and $b$ to local functions of the coordinates we will find first order constraints to be imposed in order to have a covariantly conserved stress tensor. For the $0$ component we find
\begin{eqnarray}
\nabla_\mu T^{\mu 0} &=& 0 \nonumber \\
&=&\partial_i \beta^i - 3 \partial_0 b + 3 \partial_0 b \ x \cdot x + t x_i \partial_0 \beta^i + \frac{1}{2} x_i x_j \sigma^{ij} =0
\end{eqnarray}
where $\sigma^{ij}$ is the classic shear tensor
\begin{eqnarray*}
\sigma^{ij}=\partial^i \beta^j + \partial^j\beta^i-\frac{2}{3} \delta^{ij}\partial_k \beta^k
\end{eqnarray*}
For the spatial part we find
\begin{eqnarray}
\nabla_\mu T^{\mu i} &=&0   \nonumber \\
&=&\left(\partial^i b -\partial_0 \beta^i\right) -5 x^i x^j \partial_j b - 4 t x^i \partial_0 b = 0
\end{eqnarray} 
that one again becomes the linearized form of Navier-Stokes equations (\ref{linearized_NS}) in a neighborhood of the origin.

\end{document}